# The Impact of AI-Driven Tools on Student Writing Development: A Case Study From The CGScholar AI Helper Project

[1]Raigul Zheldibayeva, [2]Ana Karina de Oliveira Nascimento, [3]Vania Castro, [3]Mary Kalantzis, and [3]Bill Cope

[1]Zhetysu University named after I. Zhansugurov; University of Illinois Urbana-Champaign - Bolashak International Scholarship, Kazakhstan
[2]Universidade Federal de Sergipe; University of Illinois Urbana-Champaign - Postdoctoral fellowship by the National Council for Scientific and Technological Development – CNPq, Brazil
[3]University of Illinois Urbana-Champaign

**Abstract**

The case study examines the impact of the CGScholar (Common Ground Scholar) AI Helper on a pilot research initiative involving the writing development of 11th-grade students in English Language Arts (ELA). CGScholar AI Helper is an evolving and innovative web-based application designed to support students in their writing tasks by providing specified AI-generated feedback. This study is one of six interventions. It involved one teacher and six students in a diverse school with low income students and explored to what extent customized AI-driven feedback can support students' writing development. The findings suggest that the implementation of AI Helper supported the development of students' writing in a number of ways. It also elicited suggestions from the teacher and students about ways of improving the still in development tool.

Key words: AI-driven writing tools, writing skills, CGScholar AI helper project, AI Feedback

## 1. Introduction

Since 2000, a group of researchers at the University of Illinois led by Bill Cope and Mary Kalantzis has been developing new online learning environments for writing assessments and feedback. The research team has developed the Common Ground Scholar or CGScholar, a social knowledge platform that supports diverse online pedagogical designs. Early in 2023, the group integrated an AI-driven review component into the writing assignments developed on the CGScholar platform, involving master's and doctoral students from the University of Illinois, mostly practicing teachers who wrote projects of around 3–5000 words including multimedia components. The core Large Language models mostly used has been successive versions of OpenAI's GPT series, integrated into CGScholar and has been designed to interface with any Learning Management System. (Kalantzis & Cope, 2025).

Peer feedback is a key feature of CGScholar's platform and pedagogic orientation. In 2024, it was decided to test the affordances of AI feedback as well. As a consequence a Retrieval-Augmented Generation (RAG) processes was created using a bounded vector database comprising 35 million tokens. This database included all graduate students' work from the past five years, as well as writings from instructors. One of the results of this bounded AI corpus was that students reported that AI reviews were surpassing their human reviews across all criteria, since human feedback was also part of the coursework procedures (Kalantzis & Cope, 2025).

Following up on this research a new version of the digital learning and teaching ecosystem called CGScholar (Cope & Kalanztis 2024) and its AI feedback component started to be designed. This version has been called CGScholar AI Helper and it aims at providing AI feedback to help high school students develop their writing. CGScholar AI Helper is a newly under development online application designed to aid students--this time focusing on high school learners--in their writing activities by providing feedback generated by artificial intelligence (AI) in a version curated by their teachers' rubrics, content focus and goals.

In order to develop the AI Helper prototype and considering the need to test the tool with real students, this paper focuses on a case study conducted at a school in the Midwest United States and is part of a larger pilot project at the University of Illinois Urbana-Champaign. This particular site involved students who were struggling writers consisting of one teacher and six students. The study was interested in exploring how AI-driven feedback could support these students' writing development. The process was based on the teachers' ordinary classroom practice and writing tasks. In this case it was a set 200-word writing assignment. This task required students to compare two reading materials selected by the teacher in accordance with curriculum objectives.

Learners were then required to interact with a curated version of AI specifically designed to meet both teachers' and students' needs and objectives. Students initially submitted their drafts through the CGScholar platform, where they received real-time feedback from an AI assistant based on the teacher's materials and rubric. After revising their work using this feedback, the assignments recieved a second-round AI feedback which could be compared to the previous one. After that, assignments were resubmitted for final evaluation by the teacher. All the feedback received was based on the teacher's customized rubric which was integrated in the AI corpus.

A qualitative approach to data analysis was employed, combining data from students' initial and revised writing assignments, selected feedback from focus groups, the teacher's post-survey, and observations from the research team. This provided additional insights into the support offered by CGScholar AI feedback, capturing both students' and teacher's perceptions of its impact.

The significance of this research relates to Cope and Kalantzis' (2025: 23) argument that "Generative AI is not suitable for unmediated use in education contexts". Therefore, recalibration is needed, when it comes to the use of AI in educational environments. For the purpose of the prototype development and the research, Generative AI was recalibrated in two ways. First, through prompt engineering. The CGScholar AI Helper follows the idea of adapting Generative AI prompts to align with what the teacher expects AI to see and provide feedback based on the specific materials provided by the teacher which were included into our ecosystem. This is done through Retrieval Augmented Generation (RAG). That way the materials added to CGScholar AI Helper are placed on top of the information available on the Generative AI tool used in the platform. Second, the ecosystem was recalibrated in terms of refinement through the incorporation of the teacher's rubric ensuring that feedback is aligned with teacher expectations. That way, when providing feedback, AI would not look for general, information related to students' writing but would focus on the teacher's materials and rubric. Therefore, far from being AI in the wild, it would mean a mediated use of Generative AI in education.

**2. Literature review**

Artificial intelligence (AI) has witnessed remarkable advancements in recent years, fostering innovation across various sectors (Shahzad et al., 2024). In educational settings, AI-powered writing tools such as Grammarly, ChatGPT, and QuillBot provide personalized grammar, style, and coherence feedback to help students improve their writing skills. While these tools offer the potential to transform traditional teaching and learning, understanding their specific impact on students' writing skills is critical.

AI-powered writing tools have shown the potential to improve student engagement and writing skills. Tools such as Grammarly provide adaptive feedback to help students improve grammar and structure (Song et al., 2023). Marzuki et al. (2024) found that these tools help students improve higher-order writing skills such as paraphrasing and clarity.

Mahapatra (2024) discusses the role of AI-driven tools in creating a positive environment for students to write. His study highlights that AI tools, in general, help create dialogue and provide immediate feedback, which can encourage students to engage more deeply in the writing process. These benefits illustrate the potential of AI tools to enhance students' learning by providing personalized support.

However, along with these benefits, AI tools have also been criticized for over-reliance and ethical concerns. Zhai et al. (2024) warn that over-reliance on AI may hinder students' development of independent writing skills. Farahani and Ghasemi (2024) also consider the potential biases in AI systems that could exacerbate inequalities if not managed properly. There are also privacy concerns, as AI tools often require access to personal data to provide tailored feedback (Bearman et al., 2022). Besides, there are problems associated with sourcing, since AI makes use of a huge database not connecting knowledge to those who produced it (Cope and Kalantzis, 2023). While these concerns highlight essential ethical and practical considerations, the potential benefits of AI in education are significant. The future of AI in education offers opportunities to support adaptable learning environments that meet changing needs (Abulibdeh et al., 2024).

Despite the negative aspects, AI offers opportunities to promote inclusion by providing personalized learning experiences. Shahzad et al. (2024) suggest that AI can bridge resource gaps and support diverse learners, while Contrino et al. (2024) highlight how adaptive learning technologies can help reduce differences in learning outcomes. This suggests that when implemented thoughtfully, AI-based tools can contribute to more inclusive educational environments by helping learners with different backgrounds and abilities to succeed.

In terms of K-12 AI research, Steinbauer et al. (2021) and Ng et al. (2021) contend that AI literacy should be integrated into the K-12 education levels, as such skills are vital for navigating life, education, and employment in an evolving society. This implies that AI literacy is also important for educators, and higher education teacher preparation programs could play a pivotal role in broadening practitioners' AI competence, equipping them to confront the demands of a dynamically evolving landscape (Southworth et al., 2023).

Addressing the research gaps in the role of AI in K-12 English Language Arts (ELA) instruction can provide valuable insights to help educators and policymakers use these tools effectively. By prioritizing the implementation of ethical AI, educational institutions can ensure that these tools support academic growth and promote inclusive and equitable learning for all students. Despite the growing adoption of AI-powered writing tools in educational contexts, limited empirical research has examined their impact on K-12 students' writing skills in ELA. Most existing studies focus on higher education institutions or regular adult learners, leaving a gap in understanding how these tools affect the development of academic writing in young students. This study aims to address this gap by assessing the impact of CGScholar AI Helper on the development of K-12 students' writing in ELA.

## 3. Methodology

*Research Design*

As mentioned previously, this case study is part of a larger research aimed at creating a prototype for the new component of CGScholar, including AI review for K-12 education. To achieve this, the research and development team adopted rapid development strategies with short, focused release cycles. This method is part of a collaborative process that is suited to research-based software development and is structured to accommodate diverse project teams (Martin, 2009; Stober & Hansmann, 2009). It emphasizes swiftly and flexibly developing software that is provided to users early on for immediate use, evaluation and feedback. This agile approach doesn't just speed up development—it fosters continuous collaboration and improvement. By engaging users early, the team can gather valuable insights and refine the software in real-time. It's a dynamic way of working that keeps projects adaptable and responsive to user needs.

That way, the research team work together with trial teachers and students in the software development process. Thus, in this paper, we attempt to not only evaluate the extent to which AI Helper supported students' writing development, but also work on the agile methodology that supports the development of the prototype. An exposition of the approach, called "cyber-social education research" (Tzirides, Saini, et al., 2023) has recently been published by the research group. As a result, with each new trial, the prototype has undergone enhancements, and the subsequent development cycle builds upon this progress. Throughout each cycle, the shared impressions of teachers and students are discussed and considered for integration. Additionally, a post-survey has been employed as part of data collection to assist the research team in refining the prototype and to understand the extent to which the CGScholar AI Helper can enhance students' writing skills and support teachers.

This study uses a qualitative approach, as qualitative research methods explore and provide a deep contextual understanding of real-world issues, including people's beliefs, perspectives, and experiences (Saunders et al., 2023). Reflexive thematic analysis was employed to evaluate the effectiveness of CGScholar AI Helper in improving 11th-grade students' English language arts writing development. The six-phase analytical process included the following steps: familiarization with the data, generating initial codes, generating themes, reviewing potential themes, defining and naming themes, producing the report (Byrne, 2021). The study design includes qualitative data analysis including coding and theming to identify the impact of AI-based feedback on student writing achievement comprehensively. Qualitative data was collected from researchers' observations, teacher post-survey feedback, students' writings including original versions and reviewed ones as well as students' feedback, provided by their participation in a focus group soon after the implementation. The research team consisted of 6 people.

*Participants*

The study took place at a public school in the Midwest region of the US, focusing on 11th-grade students. The school is located in a socioeconomical underserved area. The school has a total enrollment of 824 students for the 2024 school year, spanning grades 9 through 12. The demographic breakdown reveals that approximately 35% of the student population is white, 30% is Hispanic, and 24% identify as black. Grades 9 and 10 have the highest enrollment numbers, with 233 and 235 students respectively, while there is a decline in grades 11 (184 students) and 12 (175 students). The teacher-to-student ratio stands at 1 to 13. The school's overall dropout rate is 2.8%, predominantly affecting the Hispanic student population.

The group of students who agreed to participate are part of an English class, which has twenty-three students. To meet the protocol requirements both students and teachers were asked for permission to participate in the project. Six out of twenty-three students enrolled in the English class agreed and signed assent forms. The teacher also collected parental permission to have students participate in the experiential component of this study. To meet the language needs of parents, the consent forms were also translated into Spanish at the teacher's request. Follow-up emails and phone calls were made to ensure that the teacher and the students had submitted their consent and assent forms. Regular check-ins were conducted with participating teacher to address any concerns and provide ongoing support.

As for the participating teacher, it is a female English Language Arts (ELA) teacher who holds two master's degrees in education and is currently pursuing her doctoral studies. She was highly enthusiastic about participating in the research.

*Procedure*

The procedure began with an in-person interview with the English teacher. During the interview, she introduced her curriculum, schedule, and grading criteria. This interview allowed us to understand her grading standards, her availability to participate as well as the activity, materials and rubric she would use to assess students' writing during our implementation. Having set the dates and activities, the teacher provided the research team with her rubric and the knowledge base materials for us to feed the AI system of the platform. Once the study parameters were established, the teacher and students were trained by the research team, in separate moments, to use the CGScholar platform, with individual accounts created for each participant.

For the primary writing assignment, students completed a 200-word assignment based on the texts "The World on the Back of a Turtle" and "Returning 'Three Sisters' to Indigenous Farms Nourishes People, Land, and Cultures". Both materials were included in as knowledge base material in the CGScholar AI platform, together with the teacher's rubric, which is presented in detail in the analysis section of this paper. The assignment, chosen by the teacher, assessed students' writing in English. On the designated school day, our team was present to assist students with any technical issues on the CGScholar platform and walk them through the process. Besides, implementation day gave research team members an opportunity to make observations for further data analysis.

Initially, students submitted their first drafts directly to the platform, where the AI Helper provided real-time feedback based on the teacher's rubric. Students then revised their drafts based on this immediate AI feedback and resubmitted their assignments for one more round of AI feedback. Later, they had the chance to revise their work and submit it for final marking by the teacher. Unlike other studies focusing on feedback perception, this study specifically focuses on students' writing development and how AI Helper supports them through the iterative feedback process.

*Data collection*

Qualitative data was collected from multiple sources, including researchers' observations during the implementation, teacher feedback (post-survey), students' feedback (focus group), students' initial writing, and their updated version after AI Helper feedback. The initial interview with the teacher provided insights into her expectations and grading criteria, while additional observations provided insights into student engagement and interaction with the AI Helper tool. Further qualitative feedback from the teacher (post-survey) and students (focus group) following revision provided insights into the perceived benefits and challenges of using AI-driven feedback. More specifically, on to what extent the CGScholar AI Helper supports the development of students' writing development.

**4. Analysis**

Considering the fast development and popularization of Generative AI, one could argue that students can more easily cheat in order to complete their assignments. The argument used in this research, however, is that students learn by engaging with and utilizing machines. And this is different from having the machine complete tasks for them, which could be seen as "cheating." The point of view shared in this paper is that with proper calibration, especially in terms of prompt engineering and refinement, when used in educational contexts, Generative AI has the potential to assist students in becoming aware of how meanings are patterned. Also, it can potentially help them learn to use grammatical skills to create meaning on their own. By presenting a variety of alternative interpretations, it might also support students in developing their personal understanding of the text. In this sense, the idea of Generative AI would more likely relate to what has been called cyber-social literacy learning (Kalantzis & Cope 2025, Cope, Kalantzis & Zapata 2025).

The CGScholar AI Helper prototype has been built taking the previous thoughts into consideration. In order to do that, its features have been developed in a way to ensure that both prompt engineering and refinement are at place in the platform. In order to provide the reader with an idea of how the prototype work, firstly we present how the tool works for students and then we explain how it integrates into teachers' practice through the adoption of the teacher's prompt and rubric. We also explain how a dedicated AI corpus is aligned with teacher's work through RAG.

In order to access CGScholar AI Helper, students were directed to a website. In it, they had to login using their email and a password. For the purpose of the investigation, the research team created their logins with their school emails and a password. The login page of the tool can be seen in Figure 01.

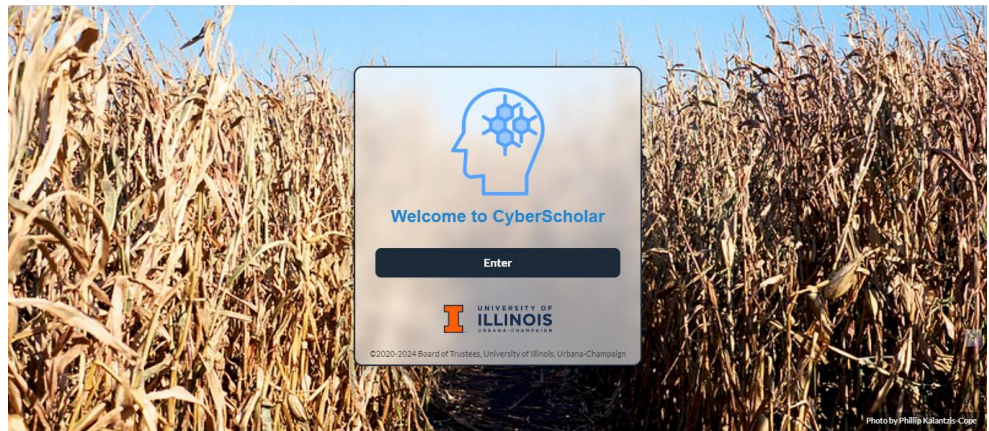
Figure 01 - Login page

After submitting their logins and passwords, students would have access to the dashboard of the tool, in which they could choose three possibilities: works, rubrics or projects. This interface can be visualized in Figure 02.

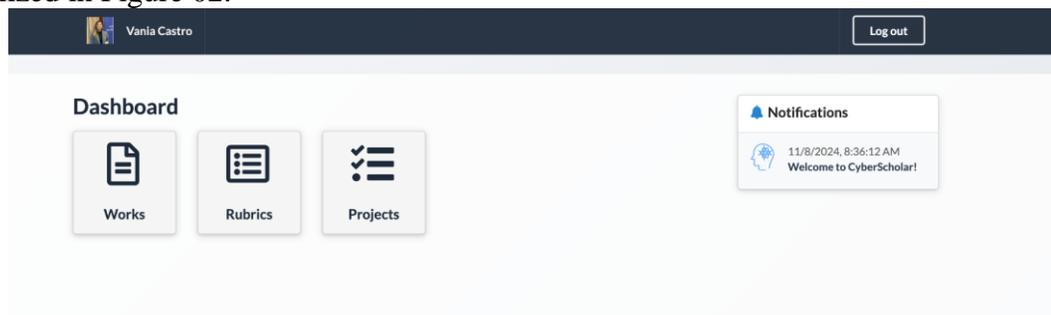
Figure 02 - CGScholar AI Helper interface

In order to have students make use of AI review, they would then need to click on 'Works'. From then, 'new works', in case it was a new one or 'existing works' if that was not students' first access. As this was all participants' initial use of the platform, they all chose 'new works', which gave them a chance to provide it with a title and a subtitle (if applicable). That would take them to an specific environment called 'document editor' in which learners had the chance to write on it and/or edit their previously written assignments. After that, when they were ready, they could run AI review by choosing AI Review. For some seconds students could visualize on their screens each item being reviewed. Each of them referred to one of the teacher's criteria as stated in the rubric shared with the research team and presented later in this section.

The way the research was conducted and the prototype developed, AI was integrated to teacher's practice. The teacher told the research team which task, among the ones to be developed, she would prefer to have the AI Helper assisting her students and when it would be carried out so that the research team could be present. In addition to that, the teacher shared with the research group, the prompt used for the activity. Thus, this prompt was included in the platform before learners could run their AI review. The prompt chosen was: "The World on the Turtle's Back" Writing Assessment - How are the Indigenous values of nature, balance, and tradition still seen today? Write a paragraph that analyzes the similarities with ONE of these values in both "The World on the Turtle's Back" translated by David Cusik and "Returning 'Three Sisters' to Indigenous Farms Nourishes People, Land, and Cultures" by Christina Gish Hill."

Still as part of the integration of AI to teacher's practice, the research team included both the "The World on the Turtle's Back" translated by David Cusik and "Returning 'Three Sisters' to Indigenous Farms Nourishes People, Land, and Cultures" by Christina Gish Hill into the CGScholar platform. That relates to one of the distinctive features of AI Helper, since we embrace the approach of customizing Generative AI prompts to align with what teachers expect the AI to interpret and to provide feedback based on specific materials supplied by the teacher and incorporated into our ecosystem. This is accomplished using RAG. By doing so, the materials added to the CGScholar AI Helper are prioritized over the information available in the Generative AI tool utilized by the platform. That is one of the recalibrations we believe to be crucial when it comes to the use of AI in educational environments.

Still focusing on the necessary recalibration and integration of AI to teacher's practice, another feature of CGScholar AI Helper is refinement. By incorporating the teacher's rubric into the ecosystem, we achieve refinement that ensures feedback aligns with the teacher's expectations. This way, when the AI provides feedback, it doesn't seek general information about student writing but focuses specifically on the teacher's materials and rubric. Therefore, rather than being unregulated or "AI in the wild" (Kalantzis & Cope 2025:23), this represents a mediated use of generative AI in education.

In the specific case of this case study, the rubric for the task involved six evaluating criteria: 1. Compare and contrast: Compare/contrast author choices, central ideas, and interpretations for two or more passages or whole texts; 2. Identify: Identify and apply proper writing conventions; 3. Compose: Compose defensible claims in both individual paragraphs and an essay as a whole; 4. Introduce and Connect: Introduce and connect evidence to claim; 5. Support Evidence: Support evidence with detailed elaboration; 6. Analyze: Analyze how Indigenous culture is represented in society. When selecting AI feedback, students then had access to it right beside their writing pieces. The feedback was displayed in two ways: a) using ratings from 0 to 4 in the form of stars, and b) through the feedback in the form of text in which the upside and downside of the writing piece were highlighted. Therefore, students had the chance to visualize their feedback graded by starts (0 to 4), followed by its explanation. This can been visualized in Figure 03.

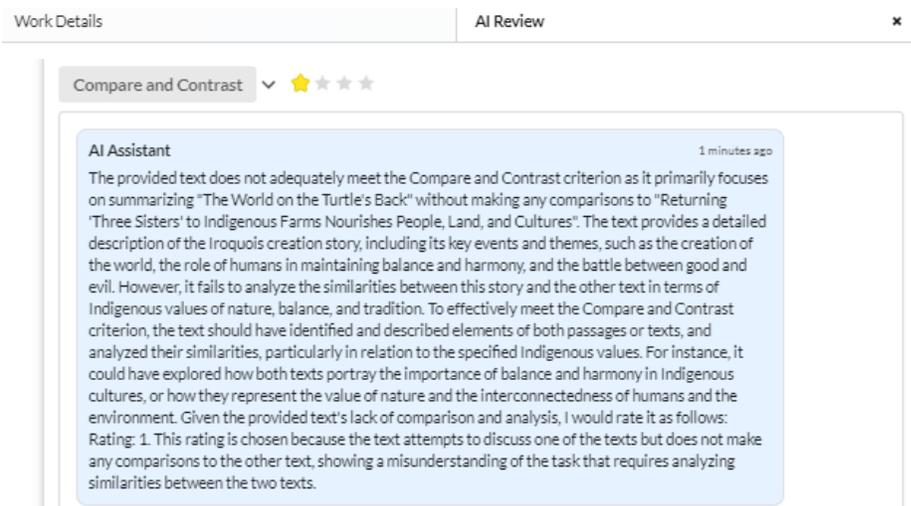

Figure 03: The AI Helper rating and feedback based on a writing sample provided by researchers

After receiving AI feedback, learners had the opportunity to read their work and revise it. Subsequently, they could run another round of AI feedback, check their improvements and then revise the text to submit it to the teacher's assessment. It is about to which extent the CGScholar AI Helper supports students' writing development that is focused the next section. More specifically, it concentrates on presenting the main findings of the first implementation of AI Helper in a high school in the US.

5. Findings

The data analysis attempts to answer the research question, which is "To what extent does CGScholar AI Helper support students' writing development?". The study involves six students in 11th grade who wrote their initial papers and then edited them accordingly after getting AI Helper feedback. The feedback was based on the teacher's provided rubric and reading materials. After the students edited their original writings, they got the second round of feedback from AI Helper, which allowed us to measure their improvements. Also, qualitative data from the focus group, teacher's post-survey, and research team observations supported this analysis. We employed reflexive thematic analysis for data analysis, creating tentative codes and later themes (Miller et.al, 2023) during and after transcribing the students' writings, students' and teacher's feedback, and research team observations. Based on the analysis, we can indicate some overall improvements: five students improved in at least one criterion, and one improved in three. One student failed to demonstrate any improvements. The aforementioned six criteria evaluated by AI helper included 1. Compare and Contrast, 2. Identify, 3. Compose, 4. Introduce and Connect, 5. Support evidence, and 6. Analyze. The rubric employed a 0-4 stars frame scale. Table 1 summarizes initial writing and edited writing performances per student for each criterion.

| Criterion | Number of students who showed improvement | Changes |
|---|---|---|
| Compare and Contrast | 3 | Two students improved from 1 to 2, one student improved from 0 to 2 |
| Identify | 0 | No improvements identified |
| Compose | 2 | Two students improved from 2 to 3 |
| Introduce and Connect | 1 | One student improved from 2 to 3 |
| Support evidence | 0 | No improvements identified |
| Analyze | 2 | Two students improved from 2 to 3 |

Table 1. Initial writing and edited writing performances.

The World on the Turtle's Back" Writing Assessment was employed to evaluate students' writings. Students were provided with two reading assignments and were expected to identify the Indigenous values of nature, balance, and tradition still seen today. The task was to write a paragraph analyzing the similarities with one of these values in both "The World on the Turtle's Back," translated by David Cusik, and "Returning 'Three Sisters' to Indigenous Farms Nourishes People, Land, and Cultures" by Christina Gish Hill.

First, we evaluated students' improvements on the Compare and Contrast criterion, as most of the students developed their writings on this criterion. Based on the completed analysis, we can state that the AI Helper supported students' writing development by providing direct feedback aligned with curriculum goals, teacher's rubric, and reading materials. For example, Student A progressed from 1 to 2, as she incorporated more detailed comparisons in her writing, but she still lacked the ability to refer to the second text. Student B developed his writing from two to three, including contextual evidence and specifying some traditions, introducing some comparisons of cultural values across the writing. The maximum improvement in the Compare and Contrast criterion was achieved by Student C, who enhanced his writing from 0 to 2 by citing some details from both required texts, successfully underlying shared values of traditions and nature. Initially, this student had showed no signs of comparison and contrast, as highlighted by the AI Helper first round of feedback: "The student does not attempt to compare or contrast the elements of two or more passages or texts. The student only focuses on explaining the significance of a tradition within a single passage without any comparison to another text or passage". After the student revised his writing based on AI Helper he could identify similarities and differences of the two texts and include some specific identification and discussion before the second round of AI feedback. So that meant that all mentioned students could demonstrate a more independent understanding of each of their text, including similarities and differences, as well as using more specific details in their revised writings.

The next criterion the students could improve in is the Compose criterion. Two students (Student B and Student D) developed their writings in composing. Both went from 2 to 3. According to the teacher's rubric, they went from "The Student attempts to compose claims, but overall the claims are too general or border on summary of the text instead of analysis" to "The Student composes defensible claims, but the claims lack a clear argument about the text specifically. Or the student requires assistance to compose a defensible claim." That means that before getting AI Helper Feedback, students could hardly compose claims and just attempted to achieve that aim, but after getting AI Helper feedback, students succeeded in composing defensible claims. AI Helper feedback could identify the lack of analytical depth and specific arguments. Moreover, it provided actionable advice to move beyond summary and focus on why the traditions are important and how they connect to overarching themes. So, for example, AI Helper suggested to Student B: "The text summarizes the ritualistic actions depicted but does not delve into an analytical argument about why this aspect is significant or how it supports an overarching thesis on tradition. The text quotes a passage and follows it with a reiteration of what the quote already states, which is more descriptive than argumentative". As we can see, the feedback directly aligns with the teacher's rubric expectation for explicit, defensible claims that make an argument about the text and respond accurately to the prompt. The student developed claims to make them more defensible. He also added some context about the importance of traditions and their continuity, by describing how traditions connect to ancestors. Moreover, according to the revised text, the student provided insights into the broader cultural significance. So that means that AI Helper feedback provided students with actionable guidance, encouraging them to refine claims and address analytical gaps, which aligned with the teacher's rubric and resulted in a higher rating.

There is another criterion in which two students (Student B and Student E) could improve from 2 to 3. AI Helper contributed to the improvements in students' performance in the Analyze criterion. After getting AI Helper feedback, both students could analyze how Indigenous culture is represented in literature and society. Both students demonstrated straightforward thematic understanding in their revised writings. For example, Student B improved his analysis by connecting the continuation of the dance rituals described in "The World on the Turtle's Back" and traditional agricultural practices in "Three sisters" to broader themes, including respect and preservation of traditions. Similarly, Student C developed her writing by adding some specific examples of human and nature kinship, underlining the animals helping to build a new world and the agricultural relationships between the "Three sisters." Their revision could lead them to understand Indigenous cosmology and agricultural symbolism better. In both cases AI Helper encouraged the students to take an action that supports the teacher's post-survey feedback, which states that "I think this has a lot of potential [...]. The one thing it did was encourage students to revise their work, which is difficult to do. "

The last criterion students could show progress in their writing is Introduce and Connect, with the student's (Student E) score increasing from 2 to 3 after AI Helper feedback. In initial writing, the student introduced evidence from both texts, "The World on the Turtle's Back" and "Returning Three Sisters"; however, there was a lack of context and weak connections between the evidence and claim related to the kinship of humans and nature in Indigenous culture. The first round of AI Helper feedback highlighted these gaps, encouraging the student to add more detailed context and provide some context or background information for evidence. AI Helper feedback says "While there is an attempt to identify similarities between the two texts, it is superficial and general, lacking the specific details and in-depth analysis" The revised work shows improvement in contextualizing the evidence and linking it to the claim, aligning the student's writing more closely to the writing task expectations.

Focused group students' post-interview analysis indicate that students found the tool **helpful** and **direct** and encouraged them to **fix** what needed improvement. This matches the higher rating results and the teacher's post-survey feedback point, stating that AI Helper **motivates** students to revise their writing, providing them with **actionable** feedback. For example, during focus group interactions, a student stated that CGScholar AI Helper "was helpful because after writing my response, I had something to go off of to fix it and it told me exactly what I needed to fix" (Student A). Another student also mentioned the following: "I think it was helpful because I could go like, go back and see what I can do right to fix it later on" (Student F). These codes that emerged from analysis indicate that a theme which emerged from the investigation is that CGScholar AI Helper supports students' writing development due to its **effective and constructive feedback**.

## 6. Conclusion

This study aimed to explore the extent to which the usage of CGScholar AI Helper as a feedback tool supports 11th-grade students' writing development. This case study research is one of the first intervention-based empirical research investigating the impact of AI-based tools on K-12 students' writing improvement. It is also the first one involving the CGScholar ecosystem. The study was conducted by the research team involved in the broader project, which is being implemented. This paper is based on the results of the first school trial implementation.

In this specific case study, qualitative data analysis was employed and the findings we observed met our expectations related to CGScholar AI Helper, proving that it is capable of supporting students' writing development. In fact, five out of six students could improve their scores at least in one criterion and one of the students was able to succeed in three criteria after getting the first round of the AI Helper feedback. The positive impact of the tool was confirmed by focus group interview, in which students had the opportunity to state that the AI Helper feedback was helpful, direct, specific, actionable and encouraging. The results point to the fact that AI Helper feedback, during this first implementation, contributed to students' writing development due to the effective and constructive feedback provided. This finding was doubled by teacher's post-survey feedback, where she stated that the tool was able to encourage the students to revise their work and motivated them to improve their writing. Both students and teachers' provided useful feedback about how the tool could be improved. The most significant issue reported was that the AI feedback provided was too long and the language used too elaborate. Taking these points into account and the fact that this is a research-based software development investigation, the research paired up with the development team have been working on adding new functions to the prototype to conform to students' and teachers' needs as realized during implementation. As a result, after this first implementation, chat boxes were included below each provided feedback. Therefore, students have the chance to have a chat with AI asking questions that relate to summarizing long-length AI Helper feedback and explanation of separate complex vocabulary. Future improvements are under development as a consequence of this pilot study, involve enabling teachers and/or students set up the size of the feedback they prefer from the CGScholar AI Helper tool. This study establishes the potential of CGScholar AI Helper as a pedagogical tool for ELA classrooms, especially in large-size writing classrooms, where students seek feedback.

One of the few limitations of the research relates to the lack of feedback on the part of some students, who did not complete pre-survey and post-survey. This fact didn't allow the research team to have a more comprehensive set of data from those two sources, so we concentrated on other sources, such as students' initial and edited writings, teacher's post-survey feedback and focus group students interview as well as research team observations. We consider that further research could be strengthened by having a more complete set of data source. We also contend that this research topic is to become dominant soon as there is a big potential for AI feedback to be integrated in classrooms. Therefore, we believe CGScholar AI Helper holds a great promise since it integrates GenAI in a customized way in order to guide students' AI use in educational contexts, taking into account teachers' design, materials, and rubric. Thus, this paper contributes to the growing knowledge of AI in education by investigating the impact of a AI-driven writing tool -- CGScholar AI Helper -- used with 11th grade English language students, providing empirical evidence to inform future applications of AI in K-12.